\documentclass[aps,pra,epsfigure,twocolumn,longbibliography,superscriptaddress]{revtex4-1}
\usepackage{dcolumn}    
\usepackage{bm} 
\usepackage{graphicx}
\usepackage{amsmath}    
\usepackage{latexsym}
\usepackage{amsfonts}   
\usepackage{amssymb}
\usepackage{float}
\usepackage{array}      
\usepackage{epsfig}
\usepackage{txfonts}
\usepackage{color}
\usepackage[colorlinks=true,linkcolor=blue,urlcolor=blue,citecolor=blue,pdfusetitle]{hyperref}
\usepackage{hyperref}

\newcommand{\ket}[1]{\left\vert#1\right\rangle}
\newcommand{\bra}[1]{\left\langle#1\right\vert}
\newcommand{\abs}[1]{\left\vert #1 \right\vert}
\newcommand{\blah}{blah\\blah\\blah\\blah\\blah. }

\usepackage{cleveref}
\usepackage{comment}

\begin{document}
\title{Work statistics and symmetry breaking in an excited state quantum phase transition}
\author{Zakaria Mzaouali}
\email{zakaria\_mzaouali@um5.ac.ma}
\affiliation{ESMaR, Faculty of Sciences, Mohammed V University in Rabat, Morocco}
\author{Ricardo Puebla}
\affiliation{Instituto de F{\'i}sica Fundamental, IFF-CSIC, Calle Serrano 113b, 28006 Madrid, Spain}
\affiliation{Centre for Theoretical Atomic,  Molecular and Optical Physics, Queen's University Belfast,  Belfast BT7 1NN, United Kingdom}
\author{John Goold}
\affiliation{School of Physics, Trinity College Dublin, College Green, Dublin 2, Ireland}
\author{Morad El Baz}
\affiliation{ESMaR, Faculty of Sciences, Mohammed V University in Rabat, Morocco}
\author{Steve Campbell}
\email{steve.campbell@ucd.ie}
\affiliation{School of Physics, University College Dublin, Belfield, Dublin 4, Ireland}
\affiliation{Centre for Quantum Engineering, Science, and Technology, University College Dublin, Belfield, Dublin 4, Ireland}

\begin{abstract}
We examine how the presence of an excited state quantum phase transition manifests in the dynamics of a many-body system subject to a sudden quench. Focusing on the Lipkin-Meshkov-Glick model initialized in the ground state of the ferromagnetic phase, we demonstrate that the work probability distribution displays non-Gaussian behavior for quenches in the vicinity of the excited state critical point. Furthermore, we show that the entropy of the diagonal ensemble is highly susceptible to critical regions, making it a robust and practical indicator of the associated spectral characteristics. We assess the role that symmetry breaking has on the ensuing dynamics, highlighting that its effect is only present for quenches beyond the critical point. Finally, we show that similar features persist when the system is initialized in an excited state and briefly explore the behavior for initial states in the paramagnetic phase. 
\end{abstract}

\date{\today}
\maketitle

\section{Introduction}
In a quantum phase transition (QPT), quantum fluctuations dominate and the system exhibits a high degree of sensitivity to the changing of an external parameter~\cite{Sachdev}. These transitions, in the ground state of a quantum system, can be classified according to the behavior of a suitable order parameter across a critical point, where first- and second-order QPTs are arguably the most prominent classes~\cite{Sachdev}. The study of such QPTs for equilibrium systems has provided deep insights into the collective properties of quantum systems, opening the possibility to exploit this critical sensitivity, for example, in protocols to achieve enhanced thermometric precision~\cite{Moha1, Moha2,mitchison2020situ} and quantum heat engines~\cite{Campisi, MossyQST}. 

While typically QPTs are exhibited in the ground state of a many-body system, certain special Hamiltonians may give rise to so-called excited state quantum phase transitions (ESQPTs)~\cite{esqpt_2020review, Leyvraz:05, Cejnar:06, Cejnar:08, Caprio:08, 2011_relano,2015_lea,2016_pra_lea,2016_prog_lea, esqpt_BE_2020, 2008_relano, 2009_relano, 2012_mazza, Brandes:13, Stransky:14, Stransky:15, Puebla:16}. Like their more traditional counterpart, ESQPTs are characterized by a similarly closing energy gap between excited states and, additionally, the density of states becomes singular around a critical excitation energy. Depending on the number of degrees of freedom,  a logarithmic divergence can be found in the density of states itself or in its higher-order derivatives~\cite{Brandes:13,Stransky:14,Stransky:15,2016_pra_lea, Caprio:08}. However, ESQPTs do not strictly occur at a fixed value of the external parameter but rather they are characterized by the lifting of degeneracy in the spectrum that occurs at progressively higher excitation energies for values of the external parameter beyond the ground state QPT~\cite{Caprio:08, 2012_mazza, 2016_pra_lea, 2011_relano}.

Studying the dynamics of a system which traverses its critical point can be broadly explored in two regimes. On the one hand traversing the critical point of a second-order QPT in a finite time leads to the emergence of critical regions, delineated by the crossover from adiabatic to impulse regimes~\cite{Zurek:05,deGrandi:10b,Polkovnikov:08,PueblaPRR,delCampoKZ}. The size of these regions are governed by the underlying critical exponents~\cite{delCampoKZ}, thus highlighting the important role that critical features in the spectrum play in dictating the dynamics. Alternatively, in order to avoid such an involved temporal analysis, the study of sudden quenches has proven to be sufficient for revealing the salient features of the effect that criticality has on the dynamics when a system is evolved across its QPT in an abrupt manner. In addition, the study of thermodynamic properties, in particular the work probability distribution and its associated moments, are readily accessible in this regime~\cite{Paraan:09,Dorner:12,prl_silva, prx_fusco, CampbellPRB, MossyNJP}. The study and analysis of how the dynamics of a system is affected by the presence of an ESQPT has only been recently explored. In Refs.~\cite{2011_relano,2015_lea,2016_pra_lea,2016_prog_lea, esqpt_BE_2020}, it was shown that the ESQPT dramatically impacts the dynamics for a suddenly quenched state, reducing the speed of the evolution due to a localization of the quantum state around the critical energy. Such impact is also visible as a cusp in the work distribution, leading to complex survival probability dynamics~\cite{2011_relano}. Remarkably, the ESQPT yields critical signatures in other quantities, such as in out-of-time correlators~\cite{Wang:19}, decoherence rates~\cite{2008_relano,2009_relano}, or in phase-space quasi-probability distributions~\cite{wang2020_husimi}, and such signatures hold under different protocols, either under infinitesimal~\cite{2017_quan} or time-dependent quenches~\cite{2020_wang_arxiv}. 
Yet, although second order QPTs and ESQPTs are intimately related to spontaneous symmetry breaking, the impact of such fundamental process in these dynamical quantities has so far been overlooked, with the notable exceptions in the realm of dynamical quantum phase transitions~\cite{Heyl:18}, where symmetry breaking upon a sudden quench is key for the emerging non-analytical behavior~\cite{2012_mazza,Puebla:13,Puebla:13b,Lang:18,Lang:18b,Zunkovic:18,halimeh2,Puebla:20}.

In this work we complement these studies by exploring the effect symmetry breaking has on the dynamics and thermodynamics of the Lipkin-Meshkov-Glick (LMG) model. 
This model has been explored extensively in the literature and in particular how underlying QPTs can affect the dynamics~\cite{LandiArXiv, LandiArXiv2, ArgentinianPRE, halimeh1, halimeh2, HeylPRB_LMG, Zunkovic:18, CampbellPRL2020, Zurek:05,deGrandi:10b,Polkovnikov:08,PueblaPRR, RMP_Silva, JoPA_Jafari, prb_najafi, prb_bose, scirep_campo, JafariPRA2018, JafariPRA2020, JafariSciRep, pra_fazio, prl_zanardi, prl_jafari, PR_Gorin, qian_pra, quan_pra, CampbellPRB, prb_silva_2014}. In addition, studies of this model have come to prominence in light of recent experimental advances in the realisation of critical systems in this class~\cite{EsslingerExp}.

The reported results are expected to apply to other similar systems such as the critical quantum Rabi model~\cite{Hwang:15,Puebla:16}. As main results, we establish that breaking the $\mathbb{Z}_2$ parity symmetry in the ferromagnetic phase leads the system to exhibit a different periodicity in its dynamics only when the quench is beyond the ESQPT. Furthermore, while the moments of the work distribution are largely unaffected by adding a small symmetry breaking term, we find the distribution itself is strongly affected. In agreement with Ref.~\cite{2011_relano, john_chapter_book}, we find that for quenches exactly to the ESQPT point the work distribution becomes non-Gaussian and further evidence of the effect of symmetry breaking can be seen in a reduction of the probability amplitudes. Furthermore, we study the entropy of the diagonal ensemble (i.e. the Shannon entropy of the work distribution) establishing that this accessible quantity~\cite{Experiment1, Experiment2, Experiment3, prb_goold_1, prb_goold_2} is very sensitive to the presence of the ESQPT. Finally, we examine the behavior for initially excited states where a qualitatively consistent behavior is found and we consider the case of states initialized in the paramagnetic phase.

\section{Model and key quantities of interest}
The Lipkin-Meshkov-Glick (LMG) model~\cite{lmg1965} describes a set of $N$ spin-$\tfrac{1}{2}$'s with infinite range interaction subject to a transverse field~\cite{prl_vidal, pre_vidal, prb_vidal, prb_castanos, quan_pra, CampbellPRB}. The fully anisotropic Hamiltonian can be written in terms of the Pauli matrices $\sigma^i_{x,z}$ acting on site $i$ as
    \begin{equation}
    \mathcal{H}=-\frac{1}{N} \sum_{i<j} \sigma_x^i \otimes \sigma_x^j  + h\sum_i \sigma_z^i, 
     \label{lmg}
    \end{equation}
where $h\geq 0$ is the strength of the magnetic field in the $z-$direction. It is convenient to recast the LMG model in terms of the total spin operators $S_\alpha\!=\!\sum_{i} \sigma_{\alpha}^i/2$, with $\alpha\!=\!\{ x,y,z\}$, 
\begin{equation}
    \mathcal{H}=-\frac{1}{N}  S_x^2  + h\: \Bigg(S_z +\frac{N}{2} \Bigg),
     \label{lmg_spin}
    \end{equation}
which can be written in a bosonic form by applying the Schwinger representation of spin operators 
\begin{align}
    S^z=t^{\dagger}t-\frac{N}{2}=\hat{n}_t-\frac{N}{2}, \: S^+=t^{\dagger}s=\big(S^-\big)^{\dagger},
\end{align}
where $S^{\pm}\!=\!S_x \pm i\:S_y$ are spin ladder operators. This results in a Hamiltonian describing a system of two species of scalar bosons $s$ and $t$, given by
\begin{equation}
    \mathcal{H}=h\:t^{\dagger}t-\frac{1}{4N}\big( t^{\dagger}s+s^{\dagger}t\big)^2.
    \label{lmg_boson}
\end{equation}
The non-zero elements of the Hamiltonian~\eqref{lmg_boson} in the basis 
\begin{equation}
    \ket{N,n_t} = \frac{\big(t^{\dagger}\big)^{n_t} \big(s^{\dagger}\big)^{N-n_t} }{\sqrt{n_t ! \big(N-n_t\big)!}} \ket{0},
    \label{base}
\end{equation}
are given by
\begin{gather}
    \bra{N,n_t}\mathcal{H} \ket{N,n_t}\!=\! h \: n_t\!-\!\frac{(n_t+1)(N-n_t)+n_t(N-n_t+1)}{4N} ,\nonumber \\
    \bra{N,n_t}\mathcal{H} \ket{N,n_t+2}\!=\!-\frac{\sqrt{(n_t+1)(N-n_t)(n_t+2)(N-n_t-1)}}{4N}, 
    \label{element_matrix}
\end{gather}
where $\ket{0}$ is the vacuum state and $0 \! \leq \! n_t \! \leq \! N $ and therefore the dimension of the Hamiltonian, Eq.~\eqref{lmg_boson}, is $N\!+\!1$. The LMG model exhibits a $\mathbb{Z}_2$ parity symmetry, given by the operator $\Pi=e^{i\pi(S_z+N/2)}=e^{i\pi t^\dagger t}$ so that $[\mathcal{H},\Pi]=0$. As a consequence, the Hamiltonian can be split into odd and even parity blocks of dimension $D_{\text{odd}}\!=\!N/2\:+1$ and $D_{\text{even}}\!=\!N/2$, respectively.

\begin{figure}[t]
    \includegraphics[width=\columnwidth]{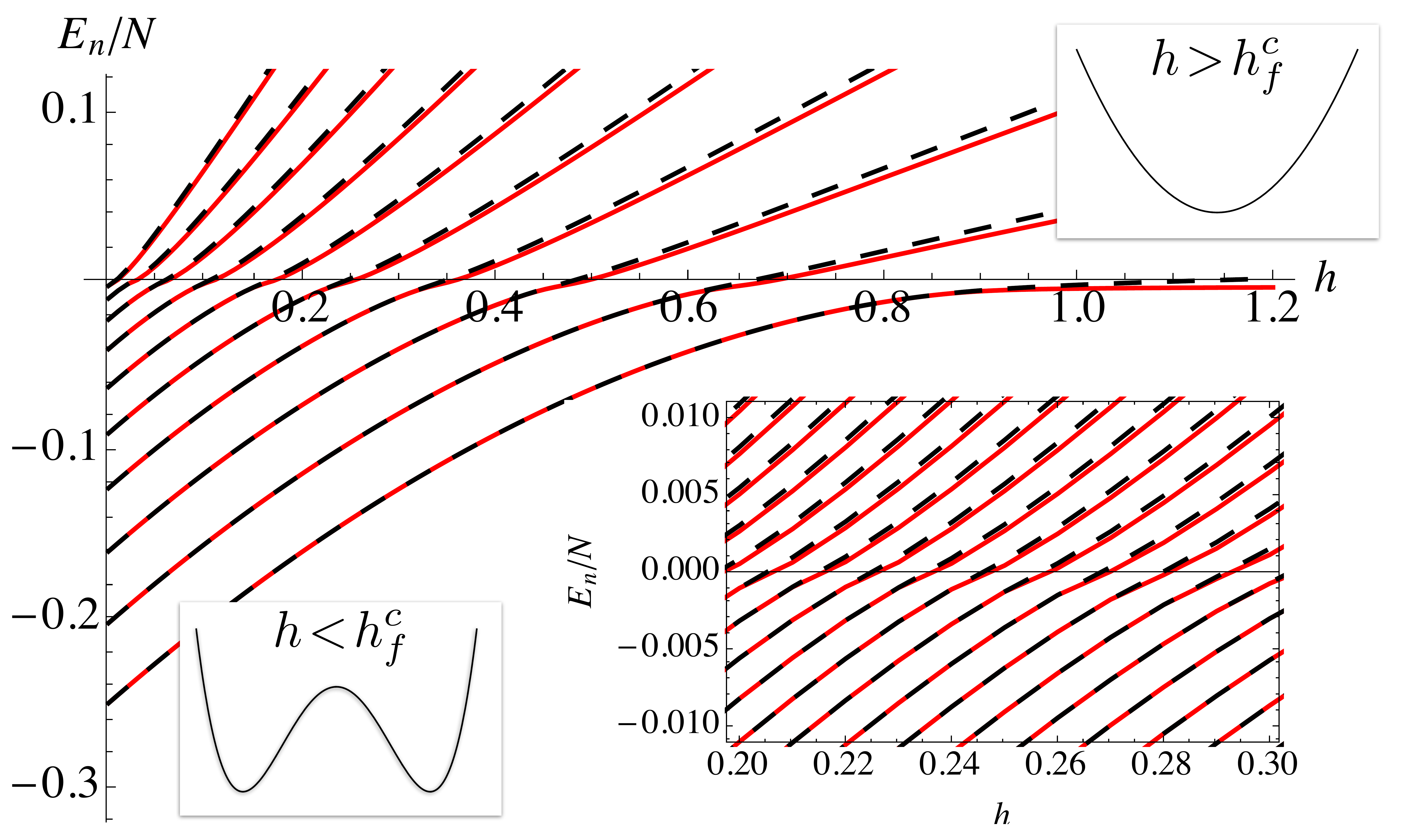}
    \caption{Spectrum of the LMG model, Eq.~\eqref{lmg_boson}, with respect to the magnetic field $h$. {\it Main:} Energy spectrum for $N\!=\!100$, showing the crossing between the critical line of the ESQPT $E_c\!=\!0$ and the lifting of the degeneracy. Only 1/5 of the total eigenstates are shown for clarity. Solid and dashed lines refer to eigenstates with opposite parity. {\it Lower right inset:} Zoom around the critical energy for $N\!=\!500$ spins. {\it Upper right+lower left insets:} Sketch of the effective potential above and below the ESQPT, respectively.}
    \label{lmg_energies}
    \end{figure}

In the thermodynamic limit, the model exhibits a spontaneous symmetry-breaking second-order QPT in the ground state at $h_c\!=\!1$~\cite{prl_vidal, pre_vidal, prb_vidal, prb_castanos} between a ferromagnetic phase ($h\!<\!1$) where the spectrum becomes doubly degenerate, and therefore the system is effectively a double well, and a paramagnetic phase ($h\!>\!1$) where all energy levels are distinct and equi-spaced. We qualitatively see the difference between these phases in Fig.~\ref{lmg_energies} where we show the eigenenergies of the LMG model as a function of $h$, Eq.~\eqref{lmg_boson} for $N\!=\!100$. The double degeneracy in the spectrum leads to another critical feature in the excited-states. For a finite value of $N$ and fixed value of $h\!<\!1$ we see that the spectrum is only doubly degenerate up to a particular energy level which characterizes ESQPT~\cite{Caprio:08}. We can identify ESQPTs either by fixing the energy while varying the control parameter of the model, or equivalently, by increasing the energy at a fixed value of the control parameter. The ESQPT refers to a non-analytical behavior of the density of states, $\nu(E)=\sum_k \delta(E-E_k)$ with $E_k$ the eigenenergies of the Hamiltonian, i.e. $\mathcal{H}=\sum_k E_k\ket{k}\bra{k}$~\cite{Caprio:08,esqpt_2020review}. As the system approaches an ESQPT in the LMG, the density of states develops a logarithmic divergence $\nu(E)\propto -\log|E-E_c|$ due to a concentration of the energy levels at $E_c\!=\!0$~\cite{prl_vidal,pre_vidal,2016_pra_lea} and for $h\!<\!1$ which is the critical region for the ESQPT in the Hamiltonian~\eqref{lmg_boson}.  In what follows we explore how dynamical signatures of the ESQPT are present in the work statistics after a sudden quench and examine the effect of breaking the $\mathbb{Z}_2$ parity symmetry.

\begin{figure*}[t]
{ \bf (a)} \hskip0.65\columnwidth {\bf (b)}\hskip0.65\columnwidth{ \bf (c)} \\
\includegraphics[width=0.65\columnwidth]{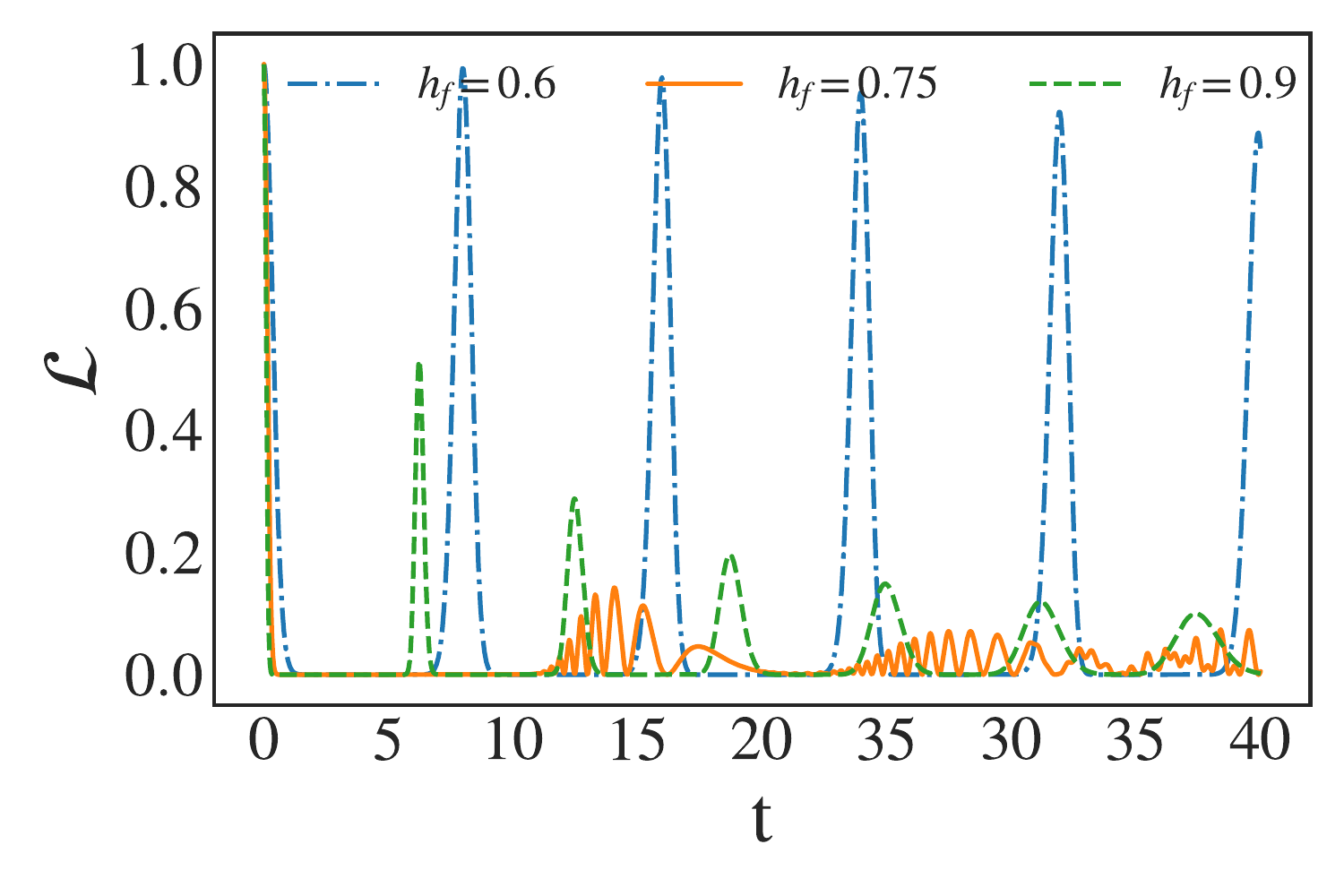}~\includegraphics[width=0.65\columnwidth]{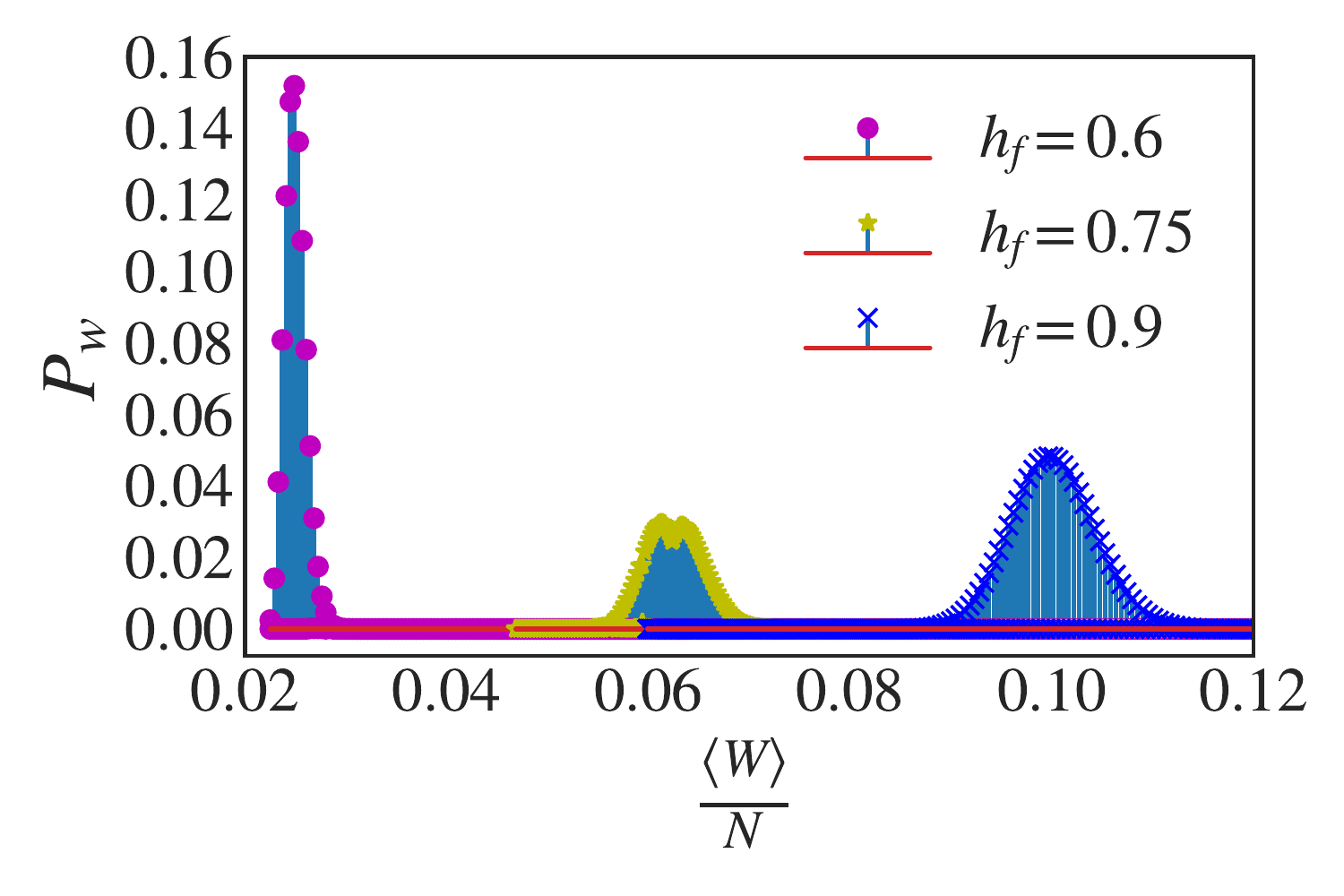}%
\includegraphics[width=0.65\columnwidth]{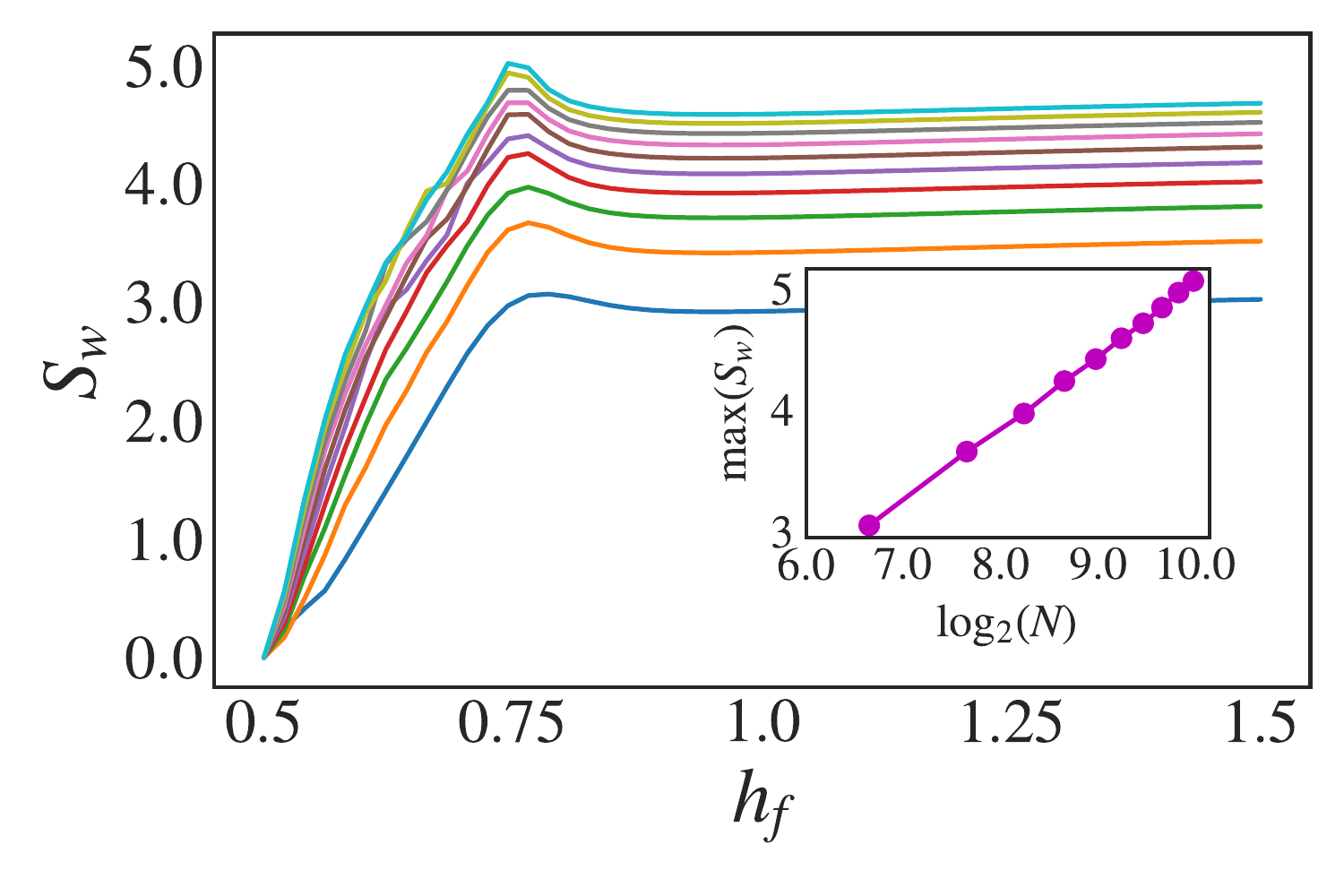}
\caption{{\bf Symmetric ground state.} (a)+(b) The survival probability, Eq.~\eqref{echo}, and the work probability distribution, Eq.~\eqref{pw}, respectively for a system size $N\!=\!2000$ and initial magnetic field $h_i=0.5$, when the quench is performed to below ($h_f\!=\!0.6$), above ($h_f\!=\!0.9$) and at the ESQPT critical point ($h_f\!=\!0.75$). (c) The Shannon entropy, Eq.~\eqref{entropy} with respect to the magnetic field $h_f$ for various system sizes $N\!=\!100 [\text{bottom,~blue}]\!\to\!1000[\text{top,~cyan}]$, the inset show the scaling of the maximum of $S_W$ with respect to $\log_2(N)$.}
\label{symb_gs}
\end{figure*}

We consider protocol where the Hamiltonian Eq.~\eqref{lmg_boson} is initialized in a particular state, $\ket{\psi_i}\!=\!\ket{\psi(0)}$, for a given value of the magnetic field, $h_i$. At $t\!\!=\!\!0$ we abruptly change the magnetic field $h_i\!\to\!h_f$ and study the time evolution of the system under the final Hamiltonian, $\mathcal{H}_f$, according to $\ket{\psi(t)}\!=\!e^{-i\mathcal{H}_f t} \ket{\psi(0)}$. In what follows, we consider $\mathbb{Z}_2$ symmetric and $\mathbb{Z}_2$  symmetry-broken ground states and excited states. A key figure of merit for studying the dynamical response of a system to such a sudden perturbation is captured by the time-dependent fidelity or survival probability which has been extensively used in studying the critical features of spin models~\cite{RMP_Silva, JoPA_Jafari, prb_najafi, prb_bose, scirep_campo, pra_fazio, prl_zanardi, prl_jafari, PR_Gorin, qian_pra, quan_pra, CampbellPRB}. Assuming the system begins in an eigenstate, it is defined as
\begin{equation}
\mathcal{L}(t)=\vert \chi(t) \vert^2
\label{echo}
\end{equation}
where
\begin{equation}
\chi(t)=\bra{\psi(0)} \psi(t) \rangle=\bra{\psi(0)} e^{-i\mathcal{H}_f t} \ket{\psi(0)}
\label{echo2}
\end{equation}
is the characteristic function of the work distribution in the case of a sudden quench and given by~\cite{prl_silva}
\begin{equation}
P_W=\sum_{m} p_{m\vert n}^\tau \delta \left( W - (E_m - E_n) \right),
\label{pw}
\end{equation}
with $E_m (E_n)$ the energy of corresponding eigenstate of the final (initial) Hamiltonian. Unless otherwise stated, we will assume the system begins in the ground state, $n\!=\!0$, and therefore $p_{m\vert 0}$ is the conditional probability of measuring $E_m$ after the quench. The moments of the work distribution due to the sudden quench can be readily determined~\cite{prx_fusco}
\begin{equation}
    \langle W^l \rangle = \sum_{m} \Big( E_m^f - E_0^i \Big)^l \abs{ \langle\psi^i_0 \vert \psi^f_m \rangle }^2\equiv (-i)^l \partial_{t}^l \chi(t) \vert_{t\to0}.
    \label{work}
\end{equation}
where the first and second moments correspond to the average work and variance, respectively. Recent proposals have demonstrated that the distribution, Eq.~\eqref{pw} is experimentally accessible~\cite{WorkDistPRL}. Under these conditions, namely initial ground state and sudden quench, the work distribution is mathematically equivalent to the infinite time average of the quantum state, i.e. the diagonal ensemble. Therefore, we have all the information necessary to determine the entropy of the diagonal ensemble~\cite{prb_goold_1, prb_goold_2, cakan2020_DE}, which is simply given by the Shannon entropy of $P_W$,
\begin{equation}
S_W=-\sum_{W}\ P_W \log_2 P_W.
\label{entropy}
\end{equation}

\section{Work statistics and symmetry breaking in ESQPTs}
Before analyzing the impact of symmetry breaking and ESQPT in the work statistics, it is convenient to find the critical value of the magnetic field $h_f^c$ for which the initially prepared ground state at $h_i$ is brought to the critical energy $E_c$ at which the ESQPT takes place. For that we rely on a semiclassical approximation, as explained in the App.~\ref{app:a}, which leads to
\begin{align}\label{eq:hfc}
h_f^c=\frac{1+h_i}{2} \qquad 0\leq h_i\leq 1.
  \end{align}
That is, for $h_f<h_f^c$ the ground state of $\mathcal{H}$ at $h_i$ is confined within the symmetry-broken phase, while a quench $h_f>h_f^c$ provides sufficient energy so that the quenched state is brought above the ESQPT where the degeneracy is lifted (cf. Fig.~\ref{lmg_energies}).

\begin{figure*}[t]
{ \bf (a)} \hskip0.65\columnwidth {\bf (b)}\hskip0.65\columnwidth{ \bf (c)} \\
\includegraphics[width=0.65\columnwidth]{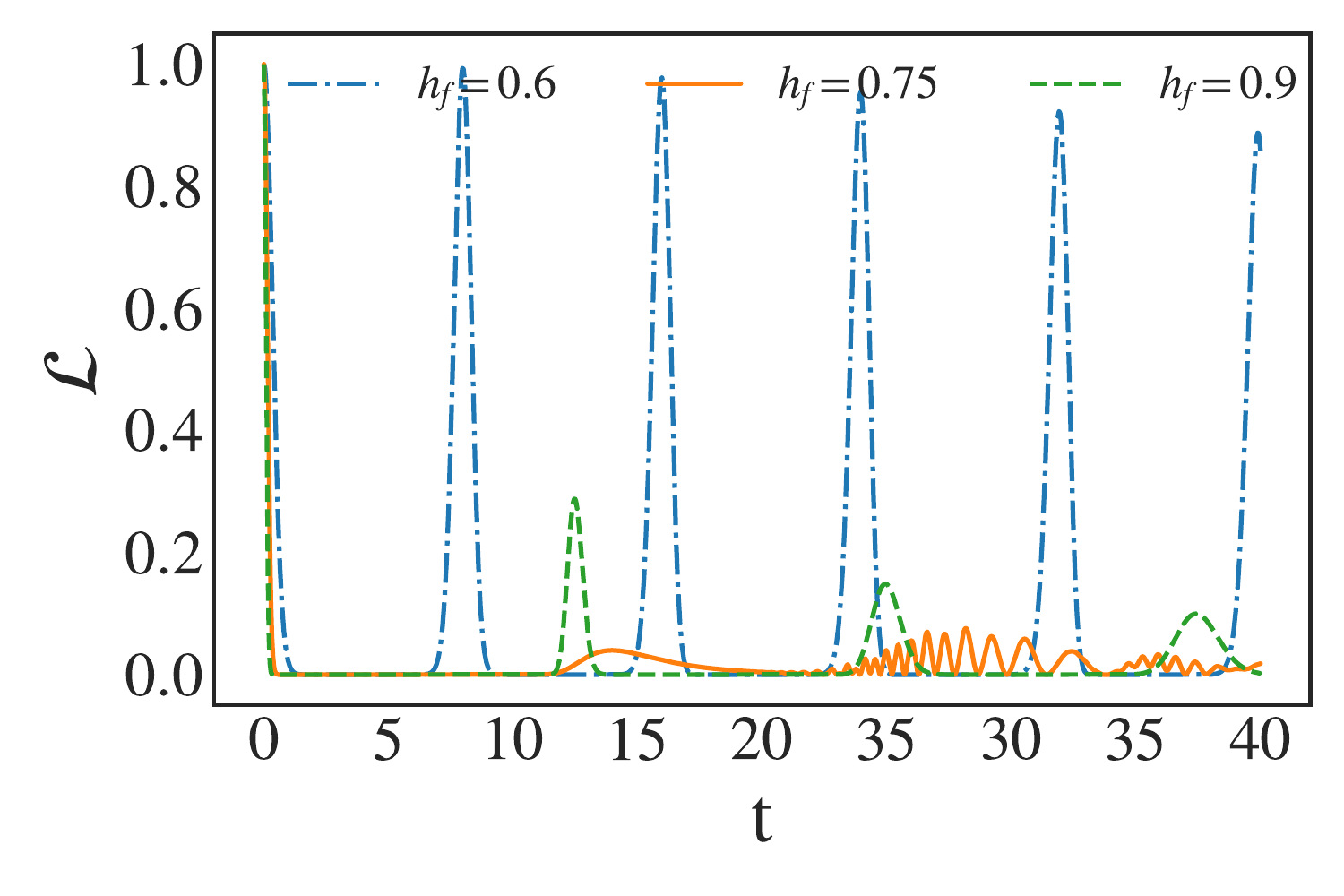}~\includegraphics[width=0.65\columnwidth]{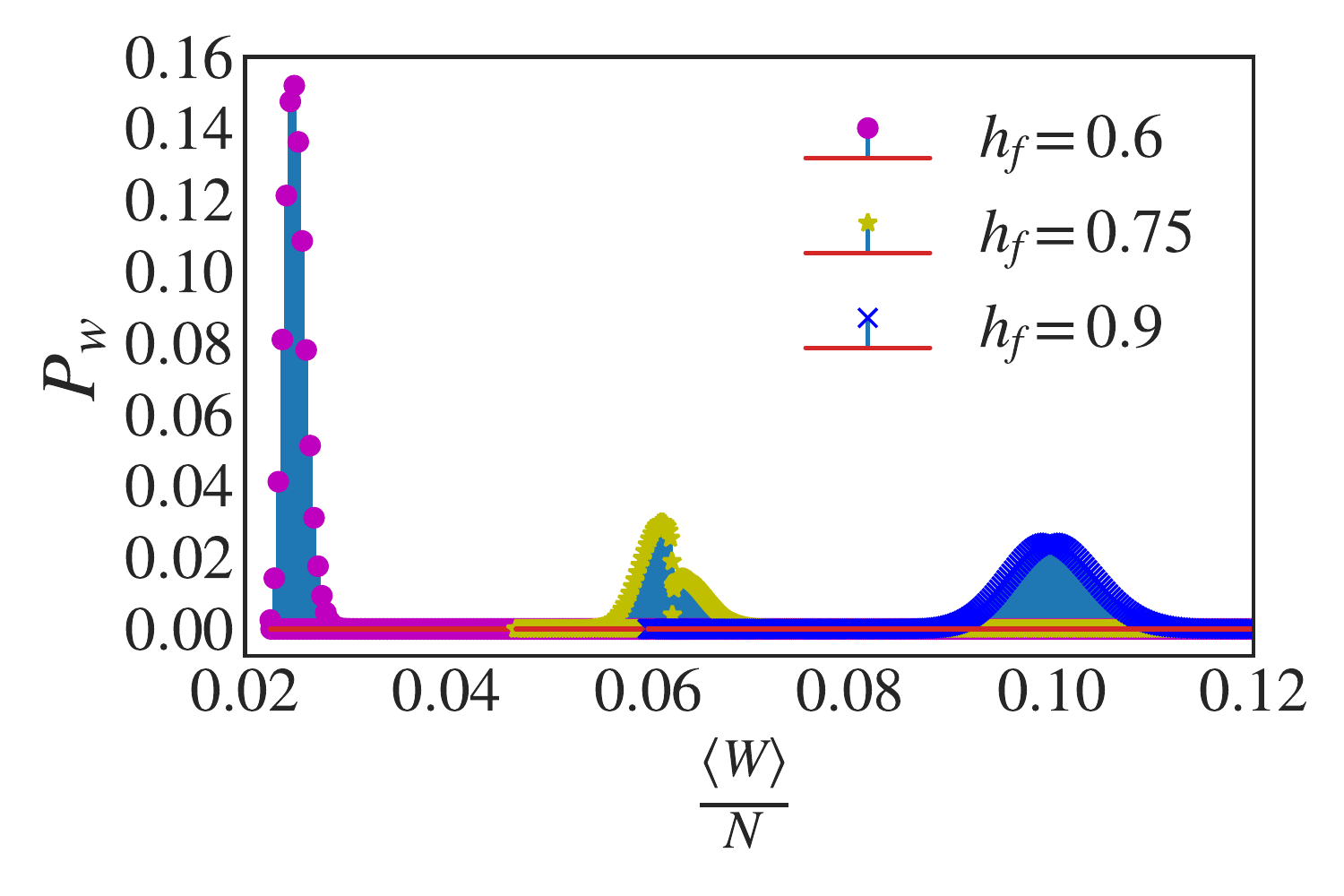}~\includegraphics[width=0.65\columnwidth]{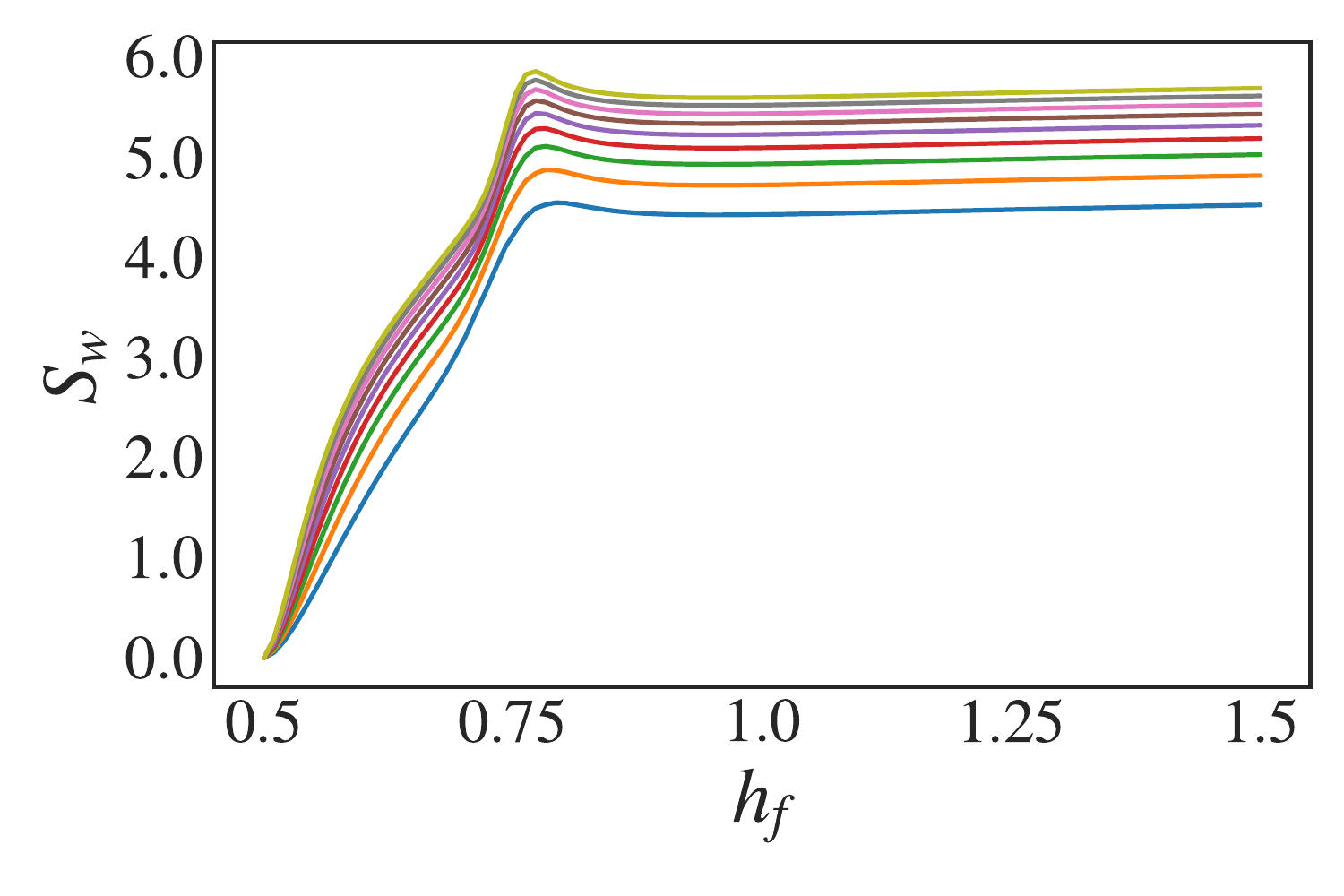}
\caption{{\bf Fully symmetry broken ground state.} (a)+(b) The survival probability, Eq.~\eqref{echo}, and the work probability distribution, Eq.~\eqref{pw}, respectively for a system size $N\!=\!2000$ and initial ground state at $h_i=0.5$, when the quench is performed to below ($h_f\!=\!0.6$), above ($h_f\!=\!0.9$) and at the ESQPT critical point ($h_f\!=\!0.75$). (c) The Shannon entropy, Eq.~\eqref{entropy} with respect to the magnetic field $h_f$ for various system sizes $N\!=\!100 [\text{bottom,~blue}]\!\to\!1000[\text{top,~yellow}]$. 
}
\label{fsb}
\end{figure*}

\subsection{Symmetric ground state}
We begin our analysis by initializing our system in the ground state of the ferromagnetic phase ($h\!<\!1$) of the LMG model, Eq.~\eqref{lmg_boson} and perform a sudden quench from $h_i\!=\!0.5$. The ESQPT corresponds to the point at which the energy of the final Hamiltonian, measured in the initial energy basis, crosses the critical line of the ESQPT $E_c\!=\!0$, cfr. Fig.~\ref{lmg_energies}. As explained above (see App.~\ref{app:a} for details), the energy of the post-quenched Hamiltonian crosses $E_c\!=\!0$  at $h_f^c\!=(1+h_i)/2=0.75$ where the ESQPT occurs.  We consider quenches to three different values of $h_f$: (i) below the ESQPT, $h_f\!<\!0.75$, (ii) to the ESQPT, $h_f\!=\! 0.75$, and (iii) beyond the ESQPT, $h_f\!>\!0.75$. We remark that qualitatively similar results hold for other choices of $h_i$ with the caveat that the location of the ESQPT is shifted accordingly as dictated by Eq.~\eqref{eq:hfc}.

Fig.~\ref{symb_gs}(a) depicts the survival proabability, Eq.~\eqref{echo} for a system size $N\!=\!2000$ and a quench starting from $h_i\!=\!0.5$. We find that quenches either sufficiently below or above the ESQPT point show qualitatively similar behaviors. In particular, for $h_f\!=\!0.6$ we find strong periodic revivals with the system almost perfectly returning to the initial state, while for $h_f\!=\!0.9$ the system still exhibits sharp revivals between periods of dynamical orthogonality, albeit with the revivals decaying in amplitude. This qualitative behavior persists for other values of $h_f$, including when quenching to and beyond the second order ground state QPT $h_c\!=\!1$~\cite{CampbellPRB} with the notable exception of in the vicinity of the ESQPT. For quenches to the ESQPT point we see the survival probability no longer exhibits such a clear periodic behavior, but instead remains dynamically close to a fully orthogonal state. The sensitivity to the presence of the ESQPT is further reflected in the work probability distribution shown in Fig.~\ref{symb_gs}(b), where we find $P_W$ is generally Gaussian for quenches to arbitrary values of $h_f$, except in the vicinity of the ESQPT, $h_f\!=\! h_f^c\!=\!0.75$, where the shape of $P_W$ changes to a double peak with the emergence of a dip, reflecting the effect of the presence of the ESQPT, as previously discussed~\cite{2011_relano}. 

We next examine the entropy of the diagonal ensemble, Eq.~\eqref{entropy}, in Fig.~\ref{symb_gs}(c) for $h_i\!=\!0.5$ as a function of the quench amplitude, $h_f$, for various system sizes. We immediately see the emergence of a peak in the entropy at the ESQPT point. We observe a logarithmic scaling of $S_W\propto \log_2(N)$ as the system size is increased, as demonstrated in the inset where we show this explicitly for the peak, however we remark that this scaling holds for any value of $h_f$. Finally, we note that the moments of the work distribution are also readily accessible, however they exhibit no sensitivity to the presence of the ESQPT, with the first (second) moments scaling linearly (quadratically) with the quench amplitude [plots not shown]~\cite{CampbellPRB}.

\subsection{Symmetry broken ground state}\label{ss:SB}
As previously mentioned, in the thermodynamic limit and for $h<1$ the LMG undergoes a spontaneous $\mathbb{Z}_2$ symmetry breaking. For finite systems, any small perturbation in $S_x$ leads to a symmetry breaking in the ferromagnetic phase ($h<1$), while it does not alter the paramagnetic phase ($h>1$). For that reason, we introduce a small perturbation $|\epsilon|\!<\!<\!1$ in $S_x$, such that it does not affect the critical features of the model, i.e.
\begin{equation}
    \mathcal{H}=-\frac{1}{N}  S_x^2  + h\: \Bigg(S_z +\frac{N}{2} \Bigg) + \epsilon\: S_x.
     \label{lmg_epsilon}
\end{equation}
In this case, the non-zero elements of the Hamiltonian in the basis, Eq.~\eqref{base}, are given by Eq.~\eqref{element_matrix} and
\begin{equation}
\bra{N,n_t}\mathcal{H} \ket{N,n_t+1}\!=\!\frac{\epsilon}{2} \sqrt{(N-n_t)(n_t+1)}.
\end{equation}
When $h\!<\!1$ the ground state of the Hamiltonian, Eq~\eqref{lmg_epsilon}, is a fully symmetry broken (FSB) ground state, i.e. a superposition of the two degenerate fully symmetric ground states with opposite parity, $\ket{\varphi_{\pm}}$ such that $\Pi\ket{\varphi_{\pm}}=\pm \ket{\varphi_{\pm}}$.  In particular, the FSB states can be written as $\ket{\varphi_{\rm FSB,\pm}}=(\ket{\varphi_{+}}\pm\ket{\varphi_-})/\sqrt{2}$ which yield a maximum value of the symmetry-breaking order parameter, $|\langle S_x \rangle|$, and are only degenerated up to an energy factor $|\epsilon|\ll 1$. 

We can now examine the effect that breaking the $\mathbb{Z}_2$ symmetry has on the figures of merit. Considering the same quench parameters as before, in Fig.~\ref{fsb}(a) we show the behavior of the survival probability. While largely consistent with the previous case, we nevertheless see some qualitative differences appearing. For quenches below the ESQPT, we see that symmetry breaking has no effect and the survival probability is identical in both instances, as can be seen by comparing the blue dot-dashed curves in Fig.~\ref{symb_gs}(a) and \ref{fsb}(a).  For quenches beyond the ESQPT, breaking the symmetry leads to a change in the period of the revivals in the survival probability. In fact, the period of these revivals double in the symmetry broken case with respect to the symmetric ground state due to the fact that the $\mathbb{Z}_2$ symmetry is no longer conserved. In this case, the absent peaks in the survival probability with respect to the symmetric ground state (cf. Fig.~\ref{symb_gs}(a)) correspond to the overlap $|\langle \varphi_{\rm FSB,-}|e^{-i\mathcal{H}_f t}|\varphi_{\rm FSB,+}\rangle|^2$, which is intimately related to the emergence of a dynamical quantum phase transition~\cite{Heyl:18,Zunkovic:18}. Quenches to the ESQPT point are notably affected by breaking the symmetry, with the system remaining closer to orthogonality throughout. 

The work distribution is similarly affected, showing evidence of the symmetry breaking only when the quench is sufficiently strong. As shown in Fig.~\ref{fsb}(b), the distribution is the same for both the symmetric and symmetry broken initial states when the quench is below the ESQPT as the energies in this region are not symmetry dependent (cf. Fig.~\ref{lmg_energies}). However, while $P_W$ retains its Gaussian profile for quenches far above/below the ESQPT point, when quenching beyond the ESQPT the distribution peak is halved, which is due to the spreading of $P_W$ over both parity subspaces of the model. The effect of symmetry breaking is most notable in the work distribution when the system is quenched to the ESQPT. Once again the distribution loses the Gaussian profile and exhibits a dip similar to symmetric case. We now find that to the left of the dip, corresponding to states below the critical energy, both the symmetric and symmetry broken initial states show the same distribution, however to the right of the dip the amplitude of the probabilities is halved again due to involvement of both parity subspaces.

The behavior of the entropy of the diagonal ensemble, Eq.~\eqref{entropy}, when we break the $\mathbb{Z}_2$ symmetry is shown in Fig.~\ref{fsb}(c) for various system sizes $N\!=\!100 [\text{blue}]\!\to\!1000[\text{yellow}] $ and is consistent with the behavior of the fully symmetric case shown in Fig.~\ref{symb_gs}(c). As in the symmetric case, $S_W$ increases quickly, peaking at the ESQPT point, $h_f^c\!=\!0.75$. We remark that, while a symmetric ground state can only populate a single parity subspace, an initial symmetry-broken ground state populates both subspaces. As a consequence, the entropy $S_W$ is larger in the symmetry-broken case by a factor $\log_2(2)=1$ when $h_f>h_f^c$, reflecting the spreading of $P_W$ over the two parity subspaces.  Finally, in contrast to these figures of merit, the first and second moments of the work distribution are unaffected by symmetry breaking.

\subsection{Weighted superposition}
\begin{figure}[t]
{ \bf (a)} \\
\includegraphics[width=0.75\columnwidth]{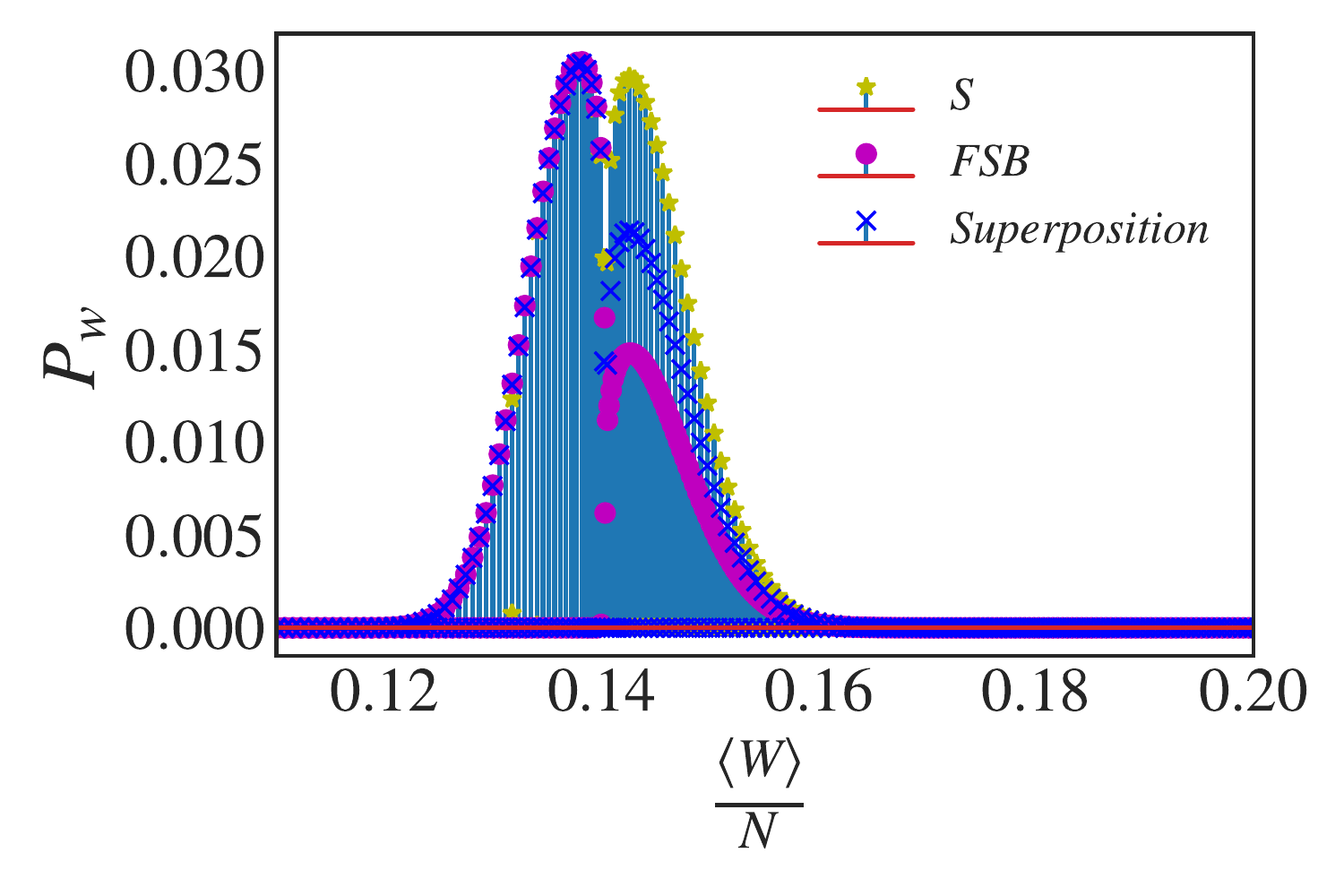}\\
{ \bf (b)} \\
\includegraphics[width=0.75\columnwidth]{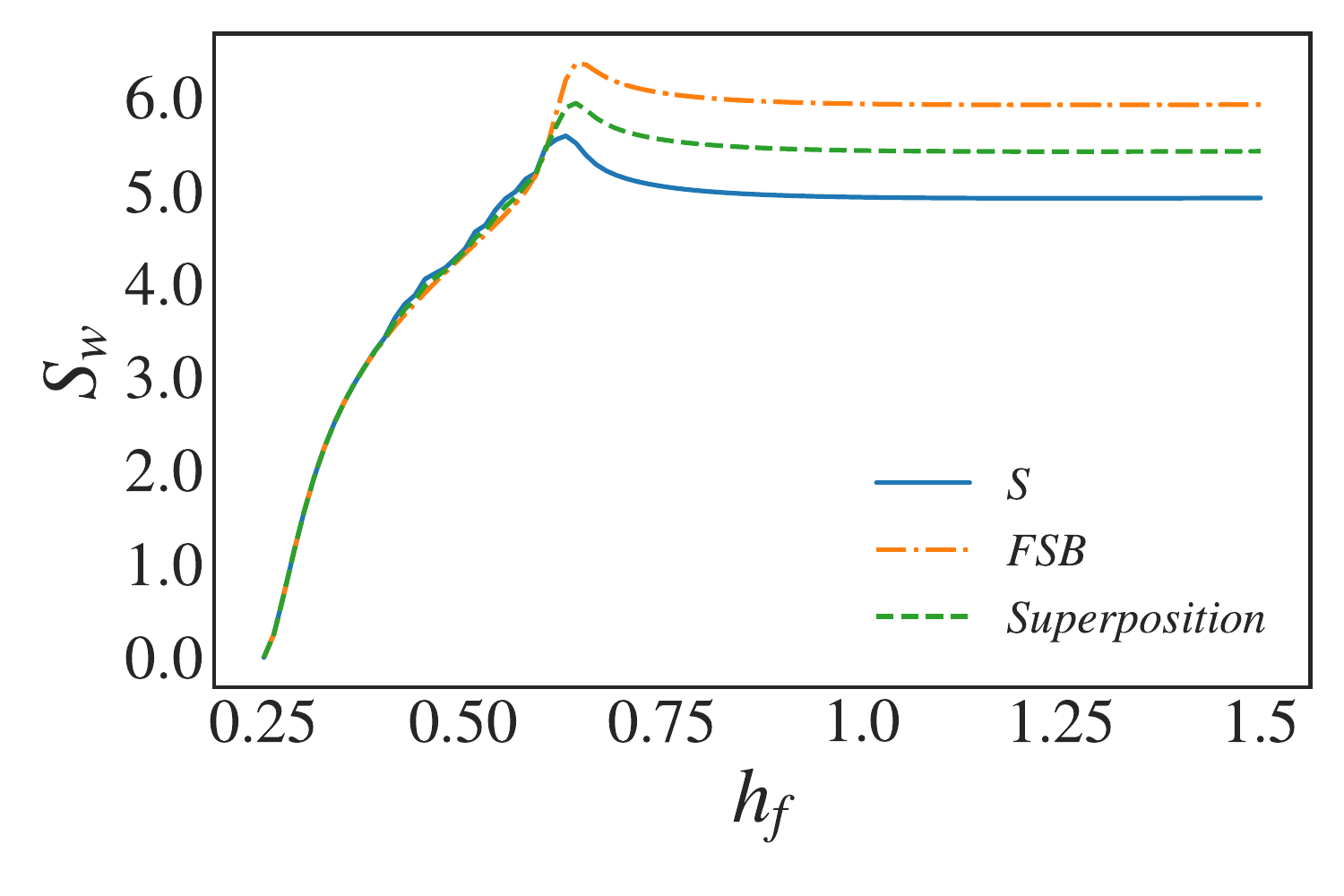}
\caption{(a) The work probability distribution $P_W$, Eq.~\eqref{pw}, and (b) The Shannon entropy $S_W$, Eq.~\eqref{entropy}, with respect to $h_f$ in the LMG model, Eq.~\eqref{lmg_epsilon}. In both panels we quench from $h_i\!=\!0.25$ in a system of size $N\!=\!1000$, initialized in the symmetric ground state (S), fully symmetry broken (FSB) ground state and in superposition between the two ground states. Note the peak at $h_f\approx 0.625$ which corresponds the critical value $h_f^c=(1+h_i)/2$ in this case (cf. Eq.~\eqref{eq:hfc}).}
\label{superposition}
\end{figure}

Having discussed the features of symmetric and fully symmetry broken ground states, we complete the picture by considering the case of an initial state which does not maximise the value of the symmetry-breaking order parameter $|\langle S_x\rangle|$ but still breaks the $\mathbb{Z}_2$ parity symmetry. As an example, we choose $|\varphi_{\rm sup}\rangle\propto 2\ket{\varphi_{+}}+\ket{\varphi_{-}}$ and we (arbitrarily) fix $h_i\!=\!0.25$ with $N\!=\!1000$. In Fig.~\ref{superposition}(a) we show the work probability distribution for all three initial states when the quench is exactly to the ESQPT critical point, $h_f^c\!=\!0.625$. All distributions exhibit the same double-peaked behavior and, furthermore, the distributions are identical to the left of the cusp. It is only for values of the work above the cusp, which corresponds to those states of the final eigenspectrum that are above the critical energy, that show the effects of symmetry breaking. Indeed, by taking a suitable superposition we can smoothly transition between the two extreme cases shown above. The entropy of the diagonal ensemble similarly reflects the effect of taking such a superposition, as shown in Fig.~\ref{superposition}(b), where $S_W$ also interpolates between the two extreme behaviors, and nevertheless clearly spotlights the presence of the ESQPT. 

\subsection{Quenching from excited states}
\begin{figure}[t]
{ \bf (a)} \\ 
\includegraphics[width=0.75\columnwidth]{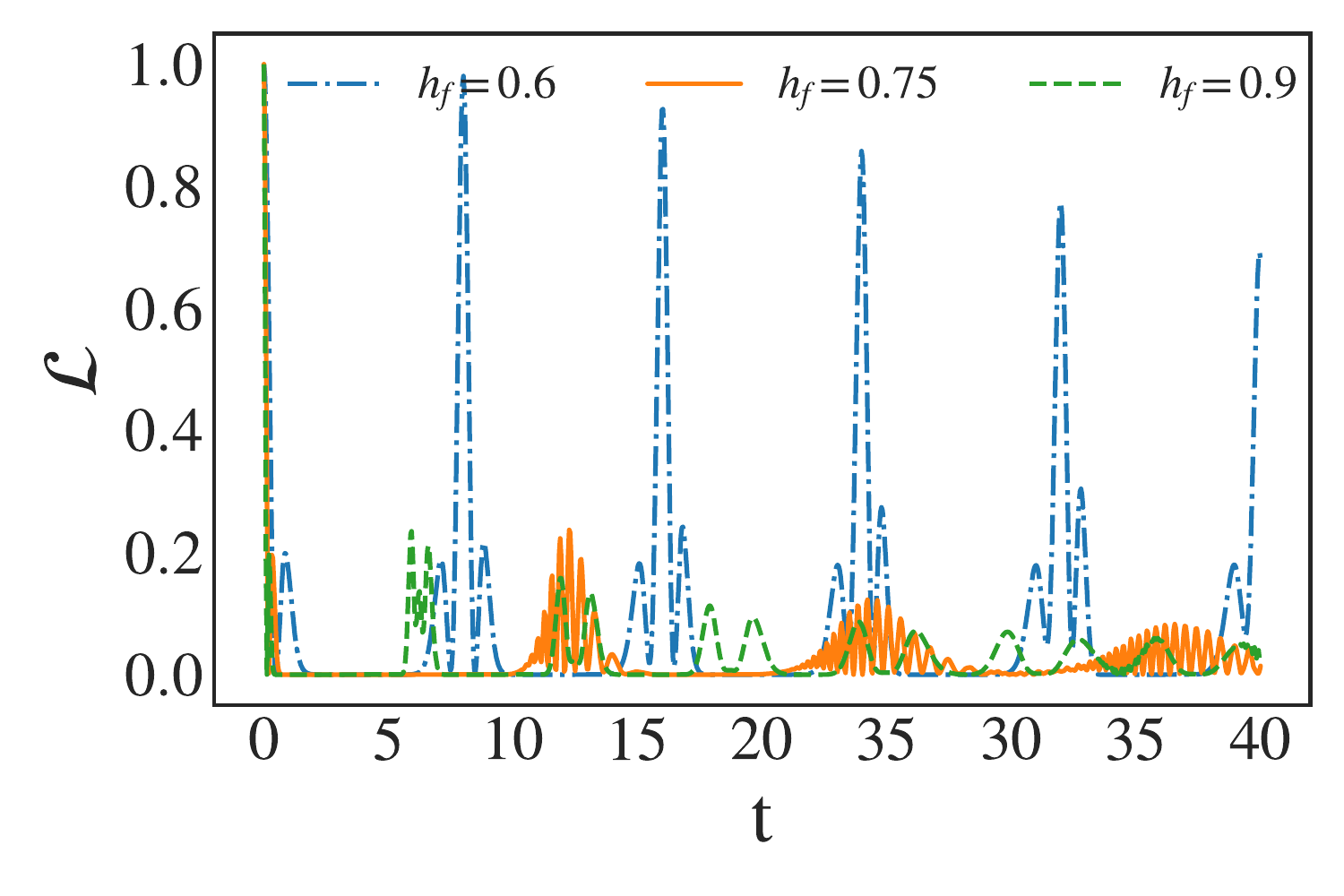}\\  {\bf (b)} \\ \includegraphics[width=0.75\columnwidth]{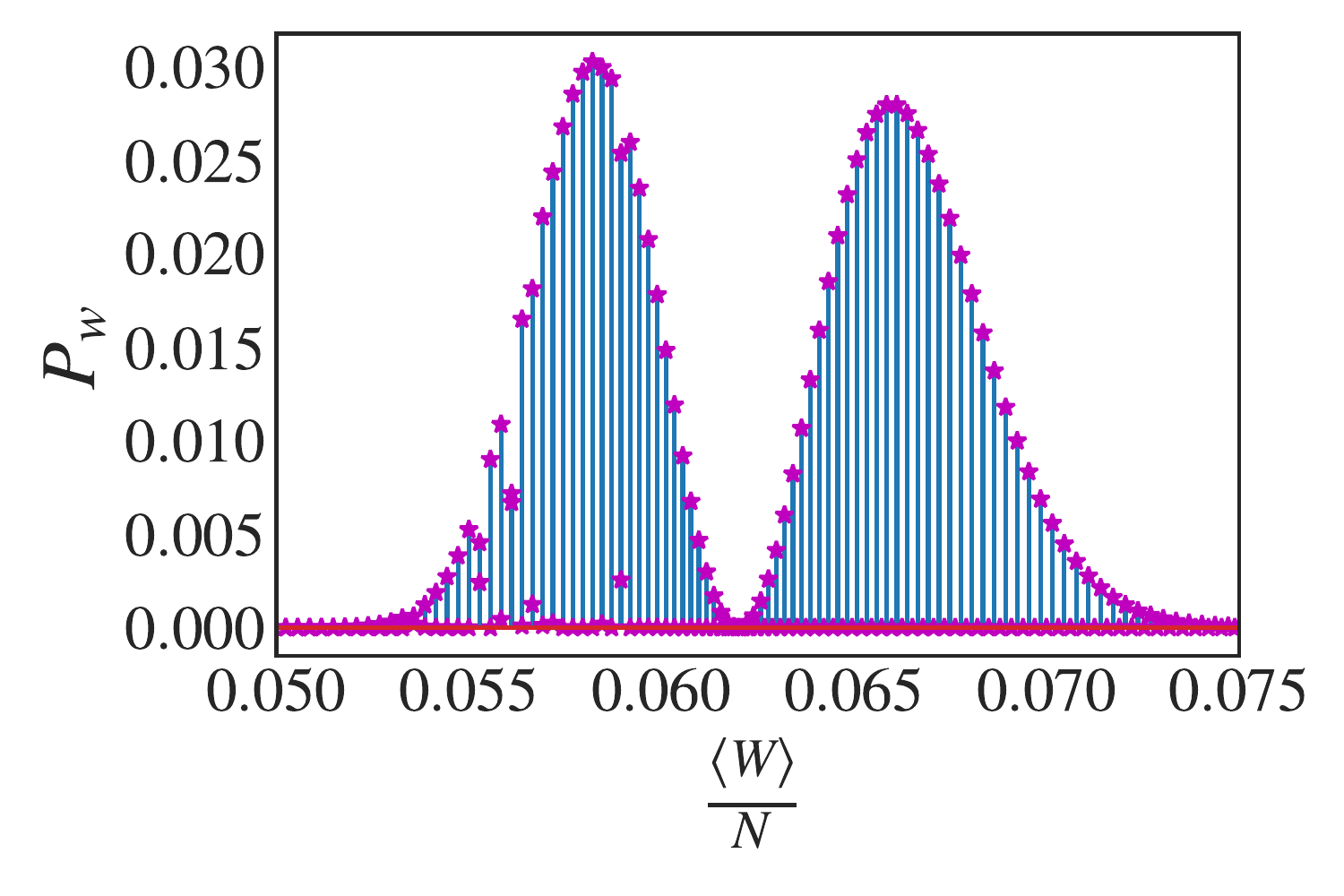}
\caption{{\bf Symmetric excited state.} (a) The survival probability, Eq.~\eqref{echo}, for a system size $N\!=\!2000$, when the quench is performed from $h_i\!=\! 0.5$ to below ($h_f\!=\!0.6$), above ($h_f\!=\!0.9$) and to the ESQPT critical point ($h_f\!=\!0.75$). (b) The work probability distribution, Eq.~\eqref{pw}, for a system of size $N\!=\!2000$ quenched from $h_i\!=\! 0.5$ to the ESQPT critical point ($h_f\!=\!0.75$).}
\label{lmg_es}
\end{figure}

An interesting feature of the model is that the double degeneracy occurring in the ferromagnetic phase is not restricted to the ground and first excited states and, in fact, extends to higher excited states up to the critical energy for $0\leq h\leq 1$. These higher excited states exhibit the same critical features and therefore here we examine whether signatures of the ESQPT are also present in the dynamics and work statistics for systems initialized in their excited state conserving the $\mathbb{Z}_2$ parity symmetry. To this end, we fix $h_i\!=\!0.5$, $N\!=\!2000$ and initialize the system in the second excited state. In Fig.~\ref{lmg_es}(a) we show the survival probability when the quench is performed to below ($h_f\!=\!0.6$), above ($h_f\!=\!0.9$) and to the ESQPT critical point $h_f^c\!=\!0.75$. Note that although $h_f^c$ corresponds to the critical value when quenching from the ground state, for low lying excited states it ensures that the quenched state has an energy in the vicinity  of the ESQPT. While the behavior is in keeping with the ground state cases, a remarkable feature emerging is the presence of higher frequencies in the revivals of the survival probability for quenches both below and above the ESQPT point and quenching to the critical energy again ensures the system remains close to orthogonality throughout the dynamics. A consequence of these higher frequencies is directly exhibited in the work probability distribution, where a bi-modal shape emerges. For quenches to the ESQPT point we find that there is still a cusp appearing in the distribution, similarly as for the ground state case. In addition, it is worth commenting that the same phenomenology of symmetry breaking applies to this scenario too.

\begin{figure*}[t]
{ \bf (a)} \hskip0.5\columnwidth {\bf (b)}\hskip0.5\columnwidth{ \bf (c)}\\
\includegraphics[width=0.65\columnwidth]{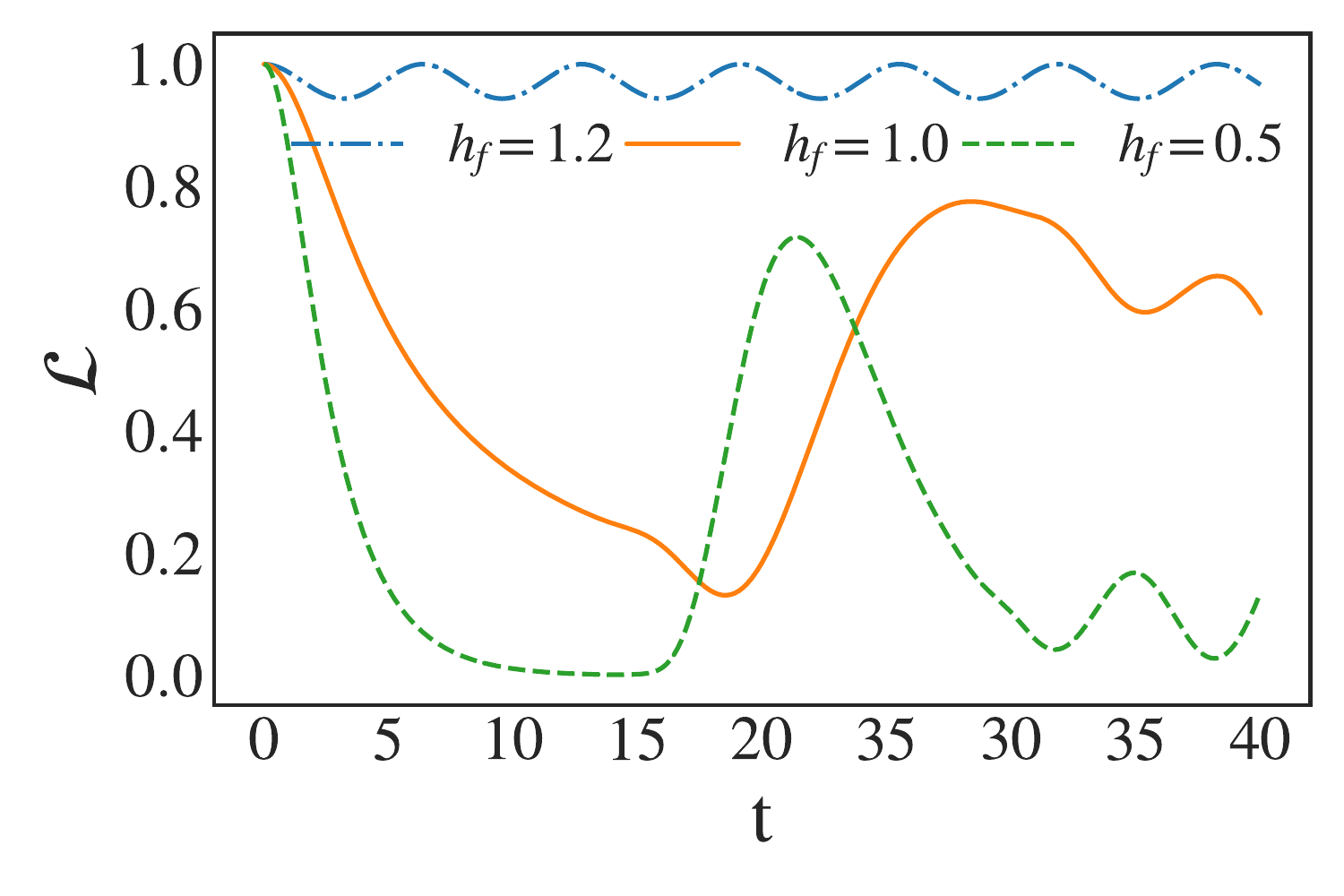}~\includegraphics[width=0.65\columnwidth]{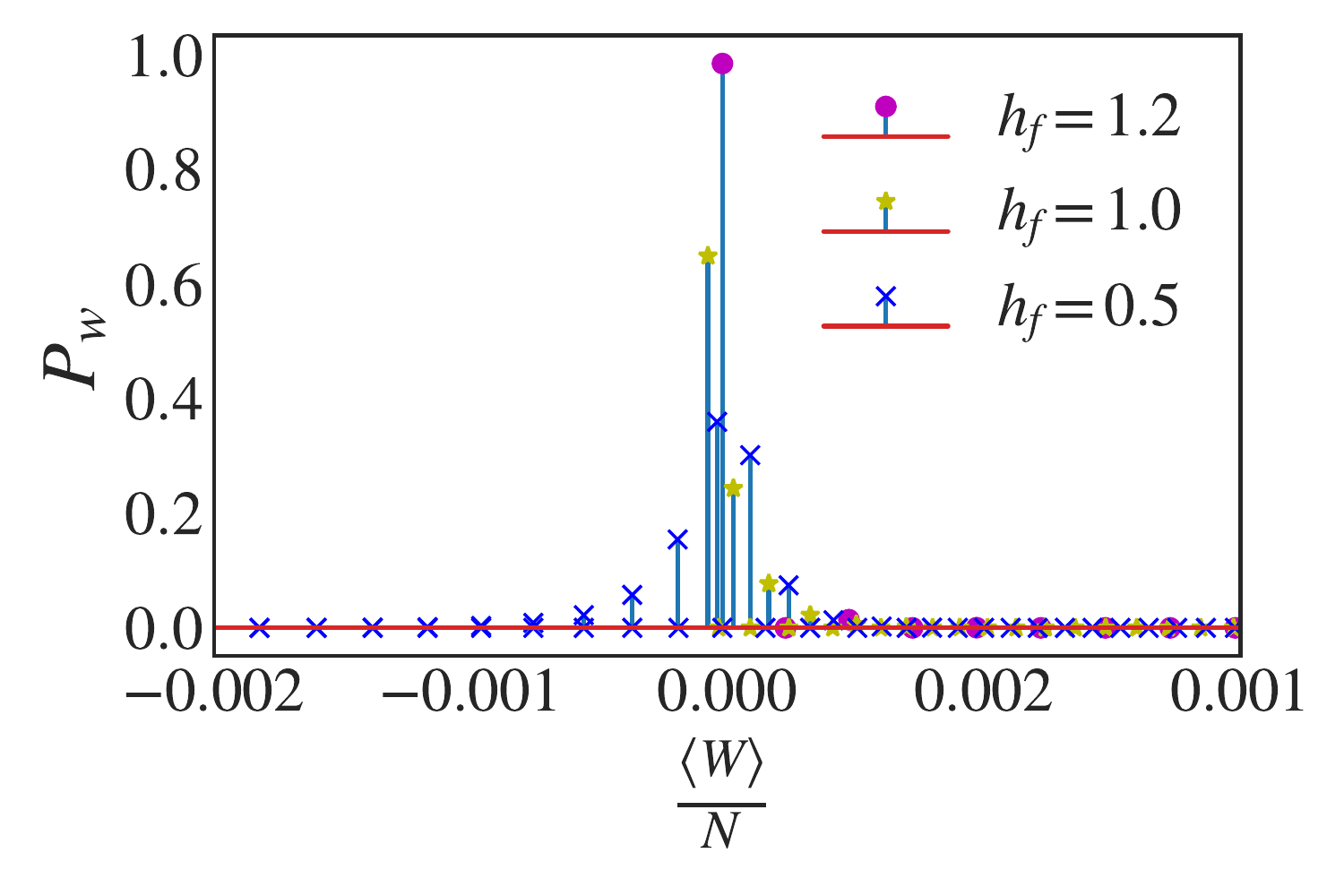}%
\includegraphics[width=0.65\columnwidth]{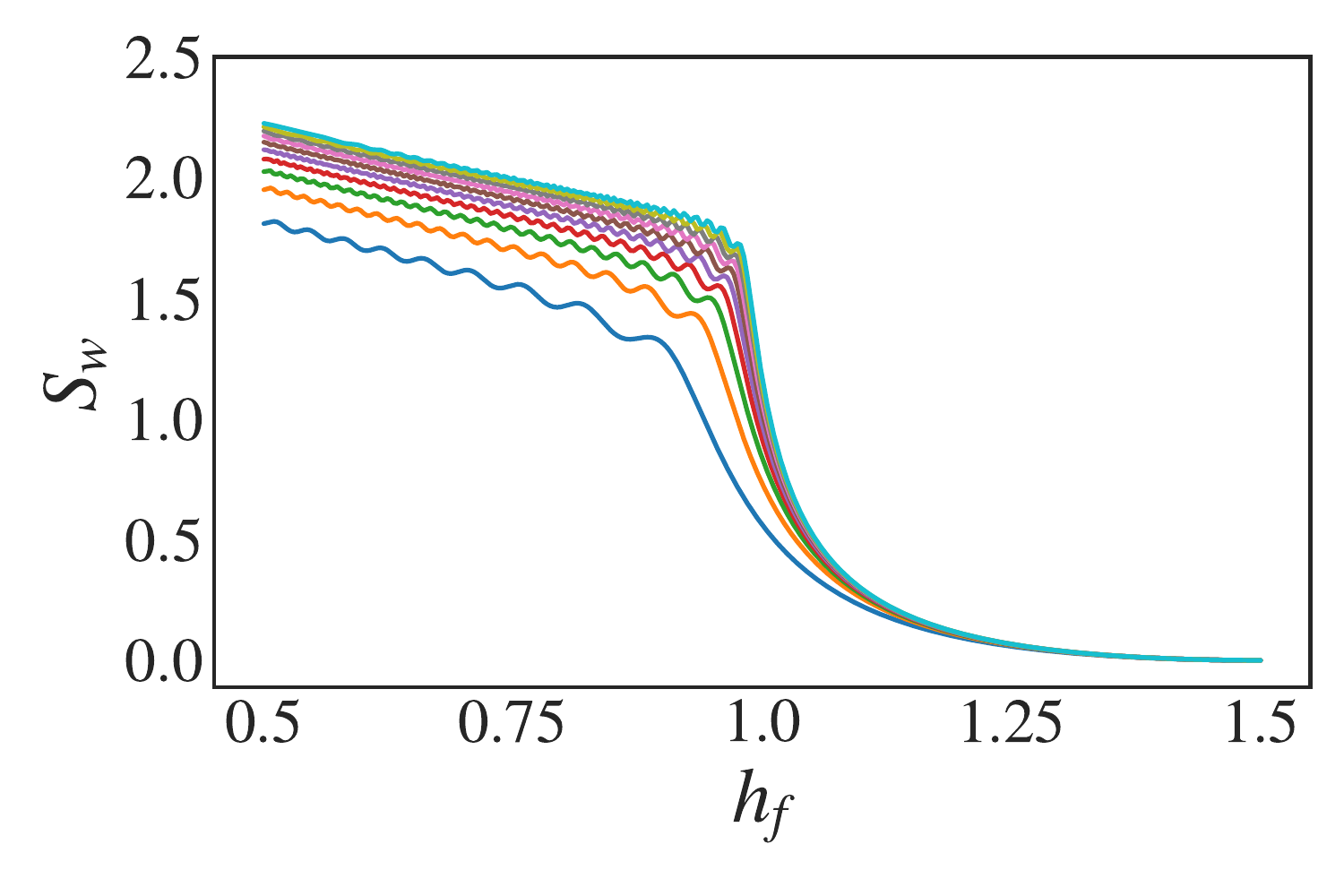}
\caption{{\bf Paramagnetic ground state.} (a)+(b) The survival probability, Eq.~\eqref{echo}, and the work probability distribution, Eq.~\eqref{pw}, respectively for a system size $N\!=\!2000$, when the quench is performed to below ($h_f\!=\!1.2$), above ($h_f\!=\!0.5$) and the second order QPT critical point ($h_f\!=\!1.0$). (c) The Shannon entropy, Eq.~\eqref{entropy} with respect to the magnetic field $h_f$ for various system sizes $N\!=\!100 [\text{bottom,~blue}]\!\to\!1000[\text{top,~cyan}]$.
}
\label{para}
\end{figure*}
\section{Quench from the paramagnetic phase}
For completeness we also consider the case of a system initialized in the paramagnetic phase ($h\!>\!1$). Unlike from the ferromagnetic phase, it is not possible to cross the ESQPT by quenching the ground state with $h_i>1$ (cf. App.~\ref{app:a}). To the contrary, the quench will be able to signal the QPT when $h_f=h_c=1$, while for $h_f<1$ the state is brought to the critical energy of the ESQPT. In Fig.~\ref{para}(a) the survival probability is shown for system size $N\!=\!2000$ and a quench starting from $h_i\!=\!1.5$. Constraining the quench to within the same phase, that is $h_f\!=\!1.2$ (blue dot-dashed curve), shows small oscillations with perfect revivals. Conversely, quenching either to the second order QPT $h_f\!=\! h_c\!=\!1$ (orange solid line) or beyond $h_f=0.5$ (dashed green line) initially drives the state far from equilibrium and the dynamics is no longer oscillatory. Furthermore, large quenches beyond the QPT drive the system to orthogonal states~\cite{CampbellPRB, CampbellPRL2020}. The work probability distribution reflects the results found for the survival probability as shown in Fig.~\ref{para}(b). $P_W$ is dominated by a single value of the work when the quench is confined within the same phase, this is a consequence of the fact that the energy levels in the paramagnetic phase are equidistant. Quenching to the QPT, we see that the distribution is still ruled by one value of the work, however other contributions are starting to emerge. A large quench crossing the QPT results in a broader probability distribution reflecting the irreversible nature of the dynamics when crossing a critical point.

Turning our attention to the entropy of the diagonal ensemble, Fig.~\ref{para}(c) shows that the entropy is small in the paramagnetic phase $h\!>\!1$, reflecting the fact that the dynamics is reversible and dominated by a single eigenstate. As we approach the QPT, the entropy sharply increases and a cusp appears tending to $h_f\!\to\!h_c=1$ as $N\!\to\!\infty$ thus indicating that the entropy of the diagonal ensemble is a faithful indicator of the ground state QPT in this case. We remark this is in contrast to the case of initial states in the ferromagnetic phase discussed previously where the presence of the ESQPT and its crossing was succinctly captured by the diagonal entropy, regardless of the presence of absence of symmetry breaking, but it was agnostic to the ground state QPT at $h_f\!=\!1$.

\section{Conclusion}
In this work we have examined the dual effect of symmetry breaking and excited state quantum phase transitions have on the dynamics of a many-body system. Focussing on the LMG model we have demonstrated that while the average work, and higher moments of the distributions, are indifferent to either the presence of an ESQPT or the effect of symmetry breaking, the distribution itself is acutely sensitive to both. Furthermore, we have established that the entropy of the diagonal ensemble is a favourable figure of merit for pinpointing and studying ESQPTs and the symmetry breaking effects~\cite{2020_wang_arxiv}. The qualitative features exhibited when the system is initialized in the ground state were shown to largely extend to initially excited states, with some notable changes, in particular, the emergence of a bimodal distribution for the work that is nevertheless sensitive to quenches to the ESQPT. Finally, we examined the behavior for quenches that start in paramagnetic phase, where the only critical features exist in the ground state and demonstrated that the entropy of the diagonal ensemble continues to be a useful tool for spotlighting the underlying critical features of the spectrum.

\acknowledgements
Z. M. thanks N. Gigena for valuable discussions. We acknowledge support from the SFI-DfE Investigator Programme (Grant No. 15/IA/2864), the European Research Council Starting Grant ODYSSEY (G. A. 758403), the SFI-Royal Society University Research Fellowship scheme, and the Science Foundation Ireland Starting Investigator Research Grant ``SpeedDemon" (No. 18/SIRG/5508)

\appendix
\section{Critical quench strength}
\label{app:a}
Here we detail how to obtain the critical value $h_f^c$ to reach the ESQPT from the ground state of $\mathcal{H}$ with $h_i$. For that we rely on a semiclassical approximation as in Ref.~\cite{prl_vidal}.  We first compute the semiclassical energy using a spin coherent representation, $\ket{\alpha}=(1+\alpha^2)^{-J}e^{\alpha S^+}\ket{J,-J}$ with $\alpha\in \mathbb{R}$ and where $\ket{J,m_J}$ denotes the standard basis of $\{S^2,S_z\}$ for the Dicke states, such that $S_z\ket{J,m_J}=m_J\ket{J,m_J}$. The energy  is
\begin{align}\label{eq:Ealpha}
E(\alpha,h)=\lim_{N\rightarrow \infty} \bra{\alpha}\mathcal{H}\ket{\alpha}=\frac{(\alpha^4-1)h-2\alpha^2}{(1+\alpha^2)^2},
\end{align}
where we have neglected the irrelevant constant energy contribution $hN/2$, which does not modify the double-well structure of $E(\alpha,h)$. 
The ground state parameter under this spin-coherent representation is achieved by minimization of $E(\alpha,h)$, which yields
\begin{align}
\alpha_{\rm gs}(h)= \begin{cases}0 \qquad \qquad h>1 \\ \pm \sqrt{\frac{1-h}{1+h}} \quad 0\leq h\leq 1 \end{cases}.
\end{align}
The solution $\alpha=0$ that ensures $dE(\alpha,h)/d\alpha=0$ becomes a local maximum for $h>1$,  which is precisely the ESQPT. For the energy functional as given in Eq.~\eqref{eq:Ealpha}, the ESQPT takes place at the critical energy $E_c=-h$ for $0\leq h\leq 1$. The two equivalent solutions for $0\leq h\leq 1$ signal the spontaneous symmetry breaking. 

In this manner, we can compute the energy of the quenched initial state as $E_q(h_i,h_f)\equiv E(\alpha_{\rm gs}(h_i),h_f)=\lim_{N\rightarrow \infty} \langle \alpha_{\rm gs}(h_i)|\mathcal{H}(h_f) |\alpha_{\rm gs}(h_i)\rangle $ where we have explicitly written the dependence of the Hamiltonian on the final parameter $h_f$. The critical quench strength follows from $E_q(h_i,h_f^c)=E_c$ which for $0\leq h_i\leq 1$ results in the simple expression given in the main text (cf. Eq.~\eqref{eq:hfc})
\begin{align}
h_f^c=\frac{1+h_i}{2}.
  \end{align}
Note however that if the $h_i>1$, then $E(h_i,h_f)=-h_f$ which corresponds to the ground state energy for $h_f\geq 1$ and the critical energy of the ESQPT for $h_f<1$. Hence, the impact of the QPT can be captured by quenching from the paramagnetic phase ($h_i>1$) to the critical point $h_f=h_c=1$. 

\bibliographystyle{apsrev4-1.bst}
\bibliography{criticalthermo}

\end{document}